\documentclass{emulateapj}
\usepackage{epsf}
\usepackage{natbib}
\usepackage{graphicx}
\usepackage{float}
\usepackage{afterpage}
 \usepackage{relsize}
\usepackage{amsmath}
\usepackage{epsfig}
\usepackage{hyperref}
\usepackage{enumitem}
\usepackage{bm,bbm,amssymb,color, amsmath, color}
\usepackage{url}

\begin{document}
\title{Radiation pressure on fluffy submicron-sized grains}
\author{Kedron Silsbee \altaffilmark{1} \& Bruce T. Draine \altaffilmark{1}}
\altaffiltext{1}{Department of Astrophysical Sciences, 
Princeton University, Ivy Lane, Princeton, NJ 08540; 
ksilsbee@astro.princeton.edu}
\begin{abstract}
We investigate the claim that the ratio $\beta$ of radiation pressure force to gravitational force on a dust grain in our solar system can substantially exceed unity for some grain sizes, provided that grain porosity is high enough.  For model grains  consisting of random aggregates of silicate spherules, we find that the maximum value of $\beta$ is almost independent of grain porosity, but for small ($<0.3 \mu$m) grains, $\beta$ actually decreases with increasing porosity.  We also investigate the effect of metallic iron and amorphous carbon inclusions in the dust grains and find that while these inclusions do increase the radiation pressure cross-section, $\beta$ remains below unity for grains with 3 pg of silicate material.  These results affect the interpretation of the grain trajectories estimated from the {\it Stardust} mission, which were modeled assuming $\beta$ values exceeding one.  We find that radiation pressure effects are not large enough for particles Orion and Hylabrook captured by {\it Stardust} to be of interstellar origin given their reported impact velocities.  We also consider the effects of solar radiation on transverse velocities and grain spin, and show that radiation pressure introduces both transverse velocities and equatorial spin velocities of several hundred meters per second for incoming interstellar grains at 2 AU.  These transverse velocities are not important for modeling trajectories, but such spin rates may result in centrifugal disruption of aggregates.  \end{abstract}
\section{introduction}
A longstanding question in the field of solar system dust grain dynamics has been the value of the ratio $\beta$ of radiation pressure force to gravitational force on a dust grain.  Because both forces scale with the inverse square of the distance to the Sun, this ratio is independent of location in the solar system.  There has been widespread agreement for several decades that as a function of grain size, $\beta$ peaks somewhere around one for grains with radii of a few tenths of a $\mu$m, but it has been a matter of debate on which side of unity the peak lies.  This is of some practical significance, since a value of $\beta$ greater than one for a certain grain size means not only the absence of orbiting grains of that size, but also a paraboloidal region of space where no interstellar grains of that size can penetrate \citep{LMG99}.  The {\it Ulysses} and {\it Galileo} missions found a deficit of grains in the mass range from 0.01 to 0.3 pg (corresponding to radii of a few tenths of a $\mu$m) inside of 4 AU in our solar system, which was interpreted as being due to radiation pressure excluding those grains from the inner solar system \citep[e.g.,][]{ Landgraf99}.  \citet{Kimura03} found that the {\it Ulysses} data are best fit assuming $\beta < 1$ for grains larger than 0.3 pg.
\par
\citet{Burns79} used Mie theory to calculate $\beta$ for spherical grains composed of a variety of materials.  They found $\beta$ values peaking around 0.6 for quartz, but well above unity for certain other materials such as graphite, iron and magnetite.  There has been speculation (e.g. \citet{Landgraf99}) that porous grains may have a significantly higher value of $\beta$.  This is intuitively appealing, since the geometric cross-section per unit mass is increased.  \citet{Saija03} studied the effect of grain structure on small (tens of nanometers) grains, and found less compact grains to have lower $\beta$ values.  We extend these computations to grain sizes as large as $a_{\rm eff} = 1 \mu$m, where $a_{\rm eff} = \left(3V/(4 \pi)\right)^{1/3}$ is the volume-equivalent radius for a grain with solid volume $V$.
\par
\citet{Kimura02} studied radiation pressure on grains formed through ballistic aggregation.  Their calculations were limited to grains with volume-equivalent radii $a_{\rm eff} \leq 0.20 \mu$m.  They used a dielectric function appropriate for obsidian, which has very little absorption.  In their calculations they found that their aggregate structures have $\beta$ values nearly an order of magnitude lower than those of a solid sphere for $a_{\rm eff} = 0.1 \mu$m, with $\beta$ dropping to very low values as $a_{\rm eff}$ shrinks below 0.1 $\mu$m.  In this paper, using a dielectric function appropriate for astrosilicates, we find more modest differences between our aggregate models and the single-sphere case, and we find that $\beta$ goes to a constant value for $a_{\rm eff} \lesssim 0.1 \mu$m, as expected when $\lambda \gg a_{\rm eff}$ and absorption dominates over scattering.
\par
The results of the {\it Stardust} mission provide new reason to pin down values of $\beta$ for larger grains with complicated geometries.  The {\it Stardust} team \citep{Westphal14} has identified three $a_{\rm eff}$ $\approx$ 0.6$\mu$m captured particles which they believe to be of interstellar origin.  The low impact speeds necessary for particles to survive the collection process require $\beta > 1$ so that solar radiation can decelerate the incoming particles prior to impact \citep{Sterken14}.  This conflicts with the results from Mie theory which show $\beta$ peaking around 0.8 for silicate grains. 
\par
 In this paper, we use accurate scattering calculations for model dust grains to examine the claim that more porous dust grains have values of $\beta$ well above those for solid spheres.   Analysis of the {\it Stardust} particles \citep{Westphal14} shows complicated compositions, including some containing ``Fe-bearing phases" said to be consistent with metallic iron.  To address this, we also calculate $\beta$ for grains with metallic iron inclusions.
 \newpage
\section{targets}
\label{sect:targets}
\begin{figure}
\centering
\includegraphics[scale = .24]{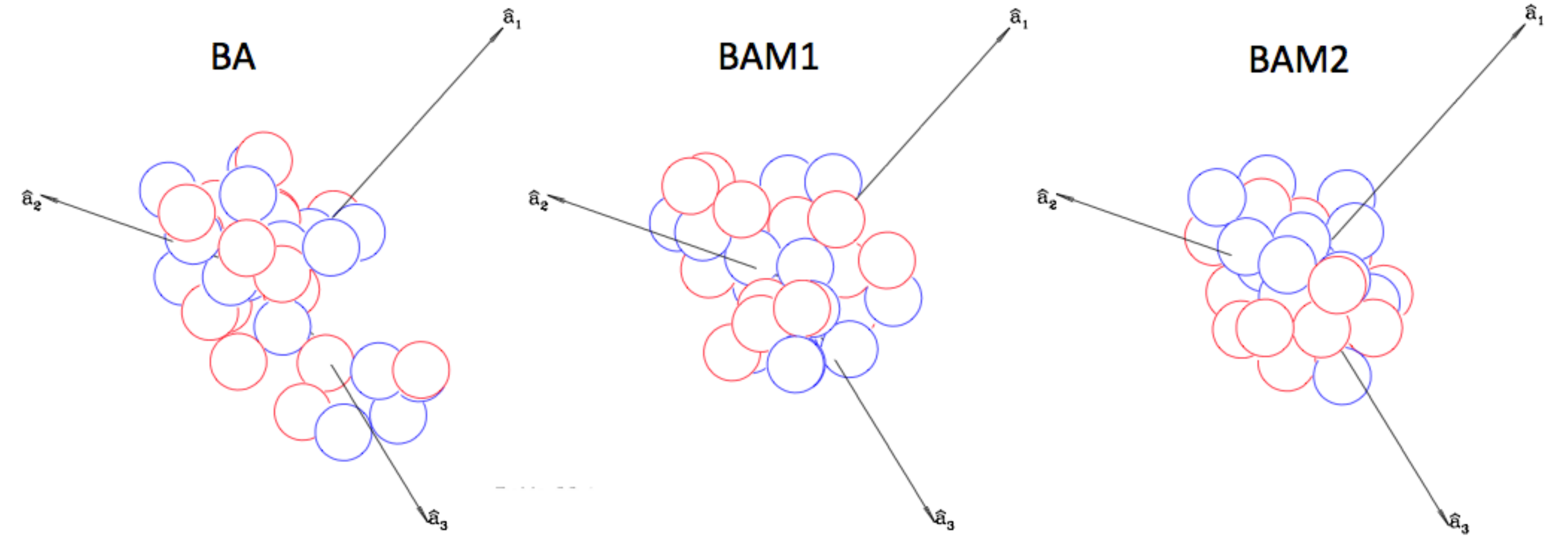}
\caption{Representatives of each class of grain for $N = 32$.  The vectors $a_i$ represent the principal axes of the grain.}
\label{grainsFigure}
\vspace{-.05cm}
\end{figure}

The grain parameters that determine $\beta$ are grain size, shape, material density, and dielectric function.  We characterize size with an effective radius $a_{\rm eff}$.  This is the radius of a solid sphere with equal mass.  For grain shapes, we use realizations of the random aggregates of \citet{Shen08}, taken from \url{http://www.astro.princeton.edu/~draine/agglom.html}.  These model grains are formed through ballistic aggregation, and Shen et al. provide three algorithms which yield grains with different porosities.  In their most porous model, denoted ``BA," a grain starts as a single sphere.  Additional spheres approach on randomly oriented trajectories, and if they impact the grain, they stick to the first point of contact.  In their intermediate density aggregates (denoted ``BAM1"), once each new sphere has impacted the target, it rolls along the surface to contact the next nearest sphere that is part of the grain.  In their most compact structures (``BAM2"),  this happens once again so each new sphere after the third ends up touching 3 spheres already part of the target before approach of the next impactor.  Shen et al. provide realizations of these structures for numbers of spheres between $N = 2^3$ and $2^{16}$.  In Figure \ref{grainsFigure} we show visualizations of one grain from each class for the realizations with 32 spheres.
\par  
 For most of these calculations we use the realizations with $N = 32$, although we verify that results are not dramatically changed by considering realizations with larger $N$.  Shen et al.\ provide average filling factors $f$ and porosities $1-f$ for these structures.  They determine $f$ by finding the uniform density ellipsoid with the same mass and moment of inertia tensor as the grain.  The filling factor $f$ is then the volume of the constituent spheres divided by the volume of this equivalent ellipsoid.  
\par
The filling factor $f$ varies substantially between the different grain models considered in this paper (from 0.15 for the BA, $N = 256$ model to 0.55 for the BAM2, $N$ = 32 model), but is fairly uniform between different realizations of models, all models having standard deviation of filling factor $\sigma_f$ on the order of $10 \%$ \citep{Shen08}.  Mean filling factors and standard deviations are presented in Table \ref{table:fillingFactors}.
\begin{table}[ht]
\caption{Filling factors}
\centering
\begin{tabular}{c c c c c}
\hline\hline
Grain Class & $N$ & $f$ & $\sigma_f$& $\sigma_f/\langle f \rangle$\\ [0.5ex] 
\hline 
BAM2 & 32 & 0.55 & 0.031 & 0.056\\ 
BAM2 & 256 &  0.42  & 0.026 & 0.061\\ 
BAM1 & 32 &  0.36 & 0.034 & 0.094\\ 
BAM1 & 256  &0.26  & 0.018 & 0.068\\ 
BA & 32  &  0.20  & 0.031& 0.155\\ 
BA & 256 & 0.15  & 0.013 & 0.086\\ [1ex]
\hline
\end{tabular}
\label{table:fillingFactors}
\end{table}
\par
For our grain composition, we use the dielectric function estimated for amorphous MgFeSiO$_4$ in \citet{Draine03}.  This silicate material is assumed to have a density $\rho_s$ = 3.8 g cm$^{-3}$, intermediate between forsterite Mg$_2$SiO$_4$ ($\rho_s$ = 3.21 g cm$^{-3}$) and fayalite Fe$_2$SiO$_4$ ($\rho_s$ = 4.39 g cm$^{-3}$).  We also consider grains with inclusions of metallic iron and amorphous carbon.  The iron inclusions have a dielectric function described in appendix B of \citet{Draine13} and density appropriate for metallic iron (7.87 g cm$^{-3}$).  We consider 3 carbon dielectric functions taken from \citet{Zubko96}, and assume a carbon density of 2.0 g cm$^{-3}$ \citep{Robertson86}.
\par
We must also specify the orientation of the grain with respect to the direction of incident radiation (i.e.\ the Sun).  This is a complicated subject, requiring averaging over a set of orientations depending on the spin state of the grain.  Fortunately, we find that orientation only affects $\beta$ at the 10\% level.  This is discussed further in Section \ref{sect:orient}.  

\section{Radiation Pressure Codes}
In this paper we use two different codes to calculate radiation pressure on dust grains.  First, we use the Multiple Sphere T Matrix (MSTM) code written by Daniel Mackowski \citep{mstmMan}.  This code calculates absorption and scattering by a collection of spheres.  The algorithm is based on decomposition of the scattered field into a superposition of vector spherical harmonics around each of the spheres \citep{Mac94, Mac96, Mac11}. The MSTM method is in principle exact, except that the expansions are necessarily truncated at a finite number of spherical harmonics for each constituent sphere (just as in Mie theory calculations for a single sphere).
\par
We also used the discrete dipole code DDSCAT \citep{Draine94, ddscat}.  It gives, among other things, the components of the mean scattered radiation field in the transverse directions.  For the radial radiation pressure, where we can do a direct comparison between the DDSCAT code and the MSTM code, we found agreement to within a few percent.  The MSTM code is not configured to easily give the transverse components of the scattered field, or to calculate torques on grains, so for this we used DDSCAT, even though it is substantially slower.

\section{calculation of $\beta$}
\label{sect:calc}
For a given grain geometry, incident wavelength $\lambda$, and dielectric function, we calculate $Q_{\rm abs}$, $Q_{\rm sca}$ and $\langle \cos{\theta_{\rm sca}} \rangle$, where $\theta_{\rm sca}$ is the angle a scattered photon makes with the direction of the incident beam, $Q_{\rm abs} = C_{\rm abs}/(\pi a_{\rm eff}^2)$, and $Q_{\rm sca} = C_{\rm sca}/(\pi a_{\rm eff}^2)$.  $C_{\rm abs}$ and $C_{\rm sca}$ are the cross-sections for absorption and scattering respectively.  The radial radiation pressure force due to sunlight is given by 
\begin{equation}
\label{Prad}
F_{\rm rad} = \int_0^\infty \frac{\pi a_{\rm eff}^2 Q_{\rm pr}(\lambda) F(\lambda)d\lambda}{c}\end{equation}
where 
\begin{equation}
Q_{\rm pr} \equiv Q_{\rm abs}+ Q_{\rm sca}(1 - \langle \cos{\theta_{\rm sca}}\rangle),
 \end{equation}
$F(\lambda)d \lambda$ is the flux of sunlight in $[\lambda, \lambda + d\lambda]$, and $c$ is the speed of light.  We approximate the Sun by a blackbody with temperature $T_{\odot} = 5777 K$ \citep{AAQ}.
 \par
 \begin{figure}
\centering
\includegraphics[scale = .5]{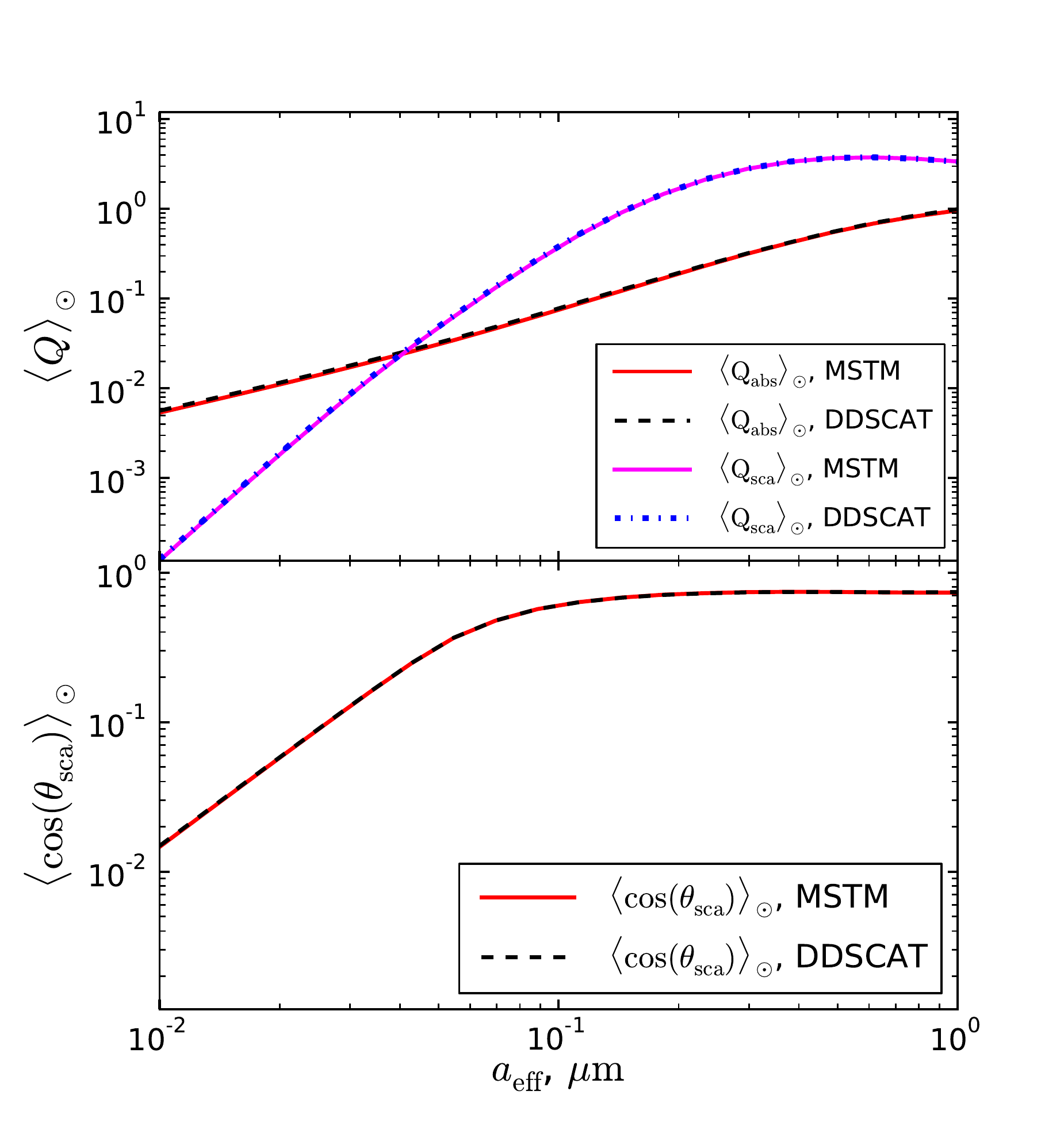}
\caption{Calculations of $\langle Q_{\rm abs}\rangle_\odot$, $\langle Q_{\rm sca} \rangle_\odot$ and $\langle \langle \cos{\theta_{\rm sca}} \rangle \rangle_\odot$ from both MSTM and DDSCAT.  This is for a particular orientation of a particular instance of a grain of the BAM1 class with 32 spheres (BAM1.32.9).  The MSTM and DDSCAT results are nearly indistinguishable.}
\label{oneGrain}
\vspace{-.05cm}
\end{figure}
\par
We define $\langle x \rangle_\odot$ as the quantity $x$ averaged over the solar spectrum.  In the case of the radiation pressure efficiency $Q_{\rm pr}$, 
\begin{equation}
\label{eq:Qpr}
\langle Q_{\rm pr} \rangle_\odot = \frac{\int_0^\infty Q_{\rm pr} F(\lambda) d\lambda}{\int_0^\infty F(\lambda) d\lambda} = \frac{\pi}{\sigma T_\odot^4} \int_0^\infty Q_{\rm pr}(\lambda) B_{\lambda} d \lambda,
\end{equation}
where $B_\lambda$ is the Planck function corresponding to $T_{\odot}$, and $\sigma$ is the Stefan-Boltzmann constant.  
\par
In Figure \ref{oneGrain}, we show a plot of $\langle Q_{\rm abs}\rangle_\odot$, $\langle Q_{\rm sca}\rangle_\odot$ and $\langle \langle \cos{\theta_{\rm sca}} \rangle \rangle_\odot$ as a function of $a_{\rm eff}$.  \kern-.5em  \footnote{See \url{http://www.astro.princeton.edu/~draine/SD2016.html}
   for additional details on computation of the results shown
   in Figure 2.}  $\langle \langle \cos{\theta_{\rm sca}} \rangle \rangle_\odot$ is the mean value of $\langle \cos{\theta_{\rm sca}} \rangle$ weighted by $F(\lambda) Q_{\rm sca}$.  We plot the results from both the MSTM code and DDSCAT.  We see that the scattering cross-section is dominant over absorption except for very small sizes.  Their relative effects on radiation pressure are more equal because of the preference for strong forward-scattering when $Q_{\rm sca} \gtrsim 1$.  
At distance $R$ from the Sun, 
\begin{equation}
\label{F}
F(\lambda) = \frac{\pi R_{\odot}^2}{R^2}B_\lambda,
\end{equation}
where $R_{\odot}$ is the radius of the Sun.  Using equations \eqref{Prad}, \eqref{eq:Qpr} and \eqref{F}, we find
\begin{equation}
F_{\rm {rad}}(a_{\rm eff}) = \frac{\pi a_{\rm eff}^2\sigma T_{\odot}^4 R_{\odot}^2}{cR^2} \langle Q_{\rm pr} \rangle_\odot.
\end{equation}
Dividing by the gravitational force on the grain
\begin{equation}
\label{eq:beta}
\beta(a_{\rm eff}) \equiv \frac{F_{\rm rad}}{F_{\rm grav}} = \frac{3 L_\odot }{16 \pi cGM_{\odot}} \frac{\langle Q_{\rm pr} \rangle_\odot}{\rho_sa_{\rm eff}},
\end{equation}
where $M_{\odot}$ is the mass of the Sun and $G$ is the gravitational constant.
To calculate $\langle Q_{\rm pr} \rangle_\odot$ numerically, we approximate $F(\lambda)$ as a sum of $N_\lambda$ delta functions, each with $\frac{1}{N_\lambda}$ of the power of the Sun, spaced such that the $i^{\rm th}$ delta function is located at $\lambda_i$ such that 
\begin{equation}
\int_0^{\lambda_i} F(\lambda) d\lambda = \sigma T_{\odot}^4 \frac{R_\odot^2}{R^2}\Big[ \frac{i-1/2}{N_\lambda}\Big], \quad i = 1, ... , N_\lambda
\end{equation}  
and
\begin{equation}
\langle Q_{\rm pr} \rangle_\odot \approx \frac{1}{N_{\lambda}} \sum_{i = 1}^{N_\lambda} Q_{\rm pr}(\lambda_i).
\end{equation}
Unless stated otherwise we use $N_\lambda = 30$ (see Appendix).  This covers the range from $\lambda_1 = 0.27 \mu$m to $\lambda_{30} = 3.34 \mu$m.
\par
\begin{figure}
\centering
\includegraphics[scale = .5]{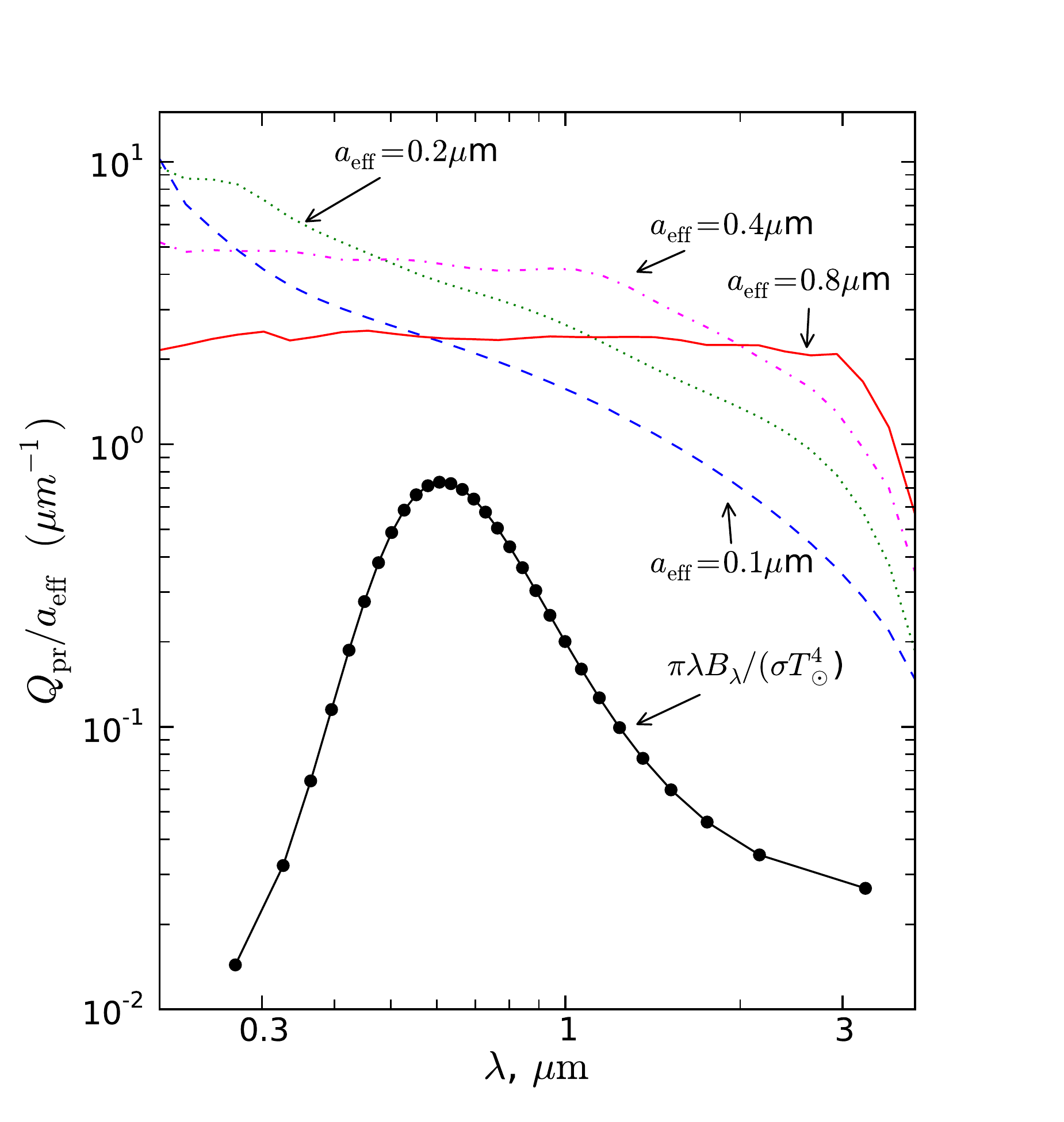}
\caption{The ratio $Q_{\rm pr} (\lambda)/a_{\rm eff}$ for different values of $a_{\rm eff}$ for a particular orientation of a particular $N$ = 32 grain of the BAM1 class (BAM1.32.1).  Also plotted is the normalized solar spectrum per unit $\ln(\lambda)$ (black curve).  The black dots are the $\lambda_i$ at which we sample the solar spectrum.  The product of these two curves is proportional to the grain acceleration per unit interval in $\ln (\lambda)$.}
\label{Qplot}
\vspace{-.05cm}
\end{figure}

In Figure \ref{Qplot}, we plot $Q_{\rm pr} (\lambda)/a_{\rm eff}$ for different values of $a_{\rm eff}$.  We normalize the $Q$'s by $a_{\rm eff}$ because acceleration due to radiation pressure is proportional to $Q_{\rm pr}/a_{\rm eff}$ (see Equation \eqref{eq:beta}). We also show the normalized solar spectrum per unit $\ln (\lambda)$.  We see that the curves of $Q_{\rm pr} (\lambda)/a_{\rm eff}$ fall off for $\lambda/a_{\rm eff} \gtrsim \pi$, and go to a constant value (proportional to 1/$a_{\rm eff}$) for $\lambda/a_{\rm eff} < 1$.  The amount of acceleration due to radiation pressure per unit $\ln(\lambda)$ is proportional to $\lambda B_\lambda Q(\lambda)/a_{\rm eff}$.  For this reason, we see that $\beta$ will be high if the curves of $Q_{\rm pr} (\lambda)/a_{\rm eff}$ are high near the peak of the solar spectrum.  This will be true for intermediate sized grains, as large ones will not have high values of $Q_{\rm pr} (\lambda)/a_{\rm eff}$ due to the $1/a_{\rm eff}$ factor and small grains will not have high values either, because of the drop-off in $Q$ for large values of $\lambda/a_{\rm eff}$.

\section{results}
\subsection{Pure Silicate Grains}
\begin{figure}
\centering
\includegraphics[scale = .5]{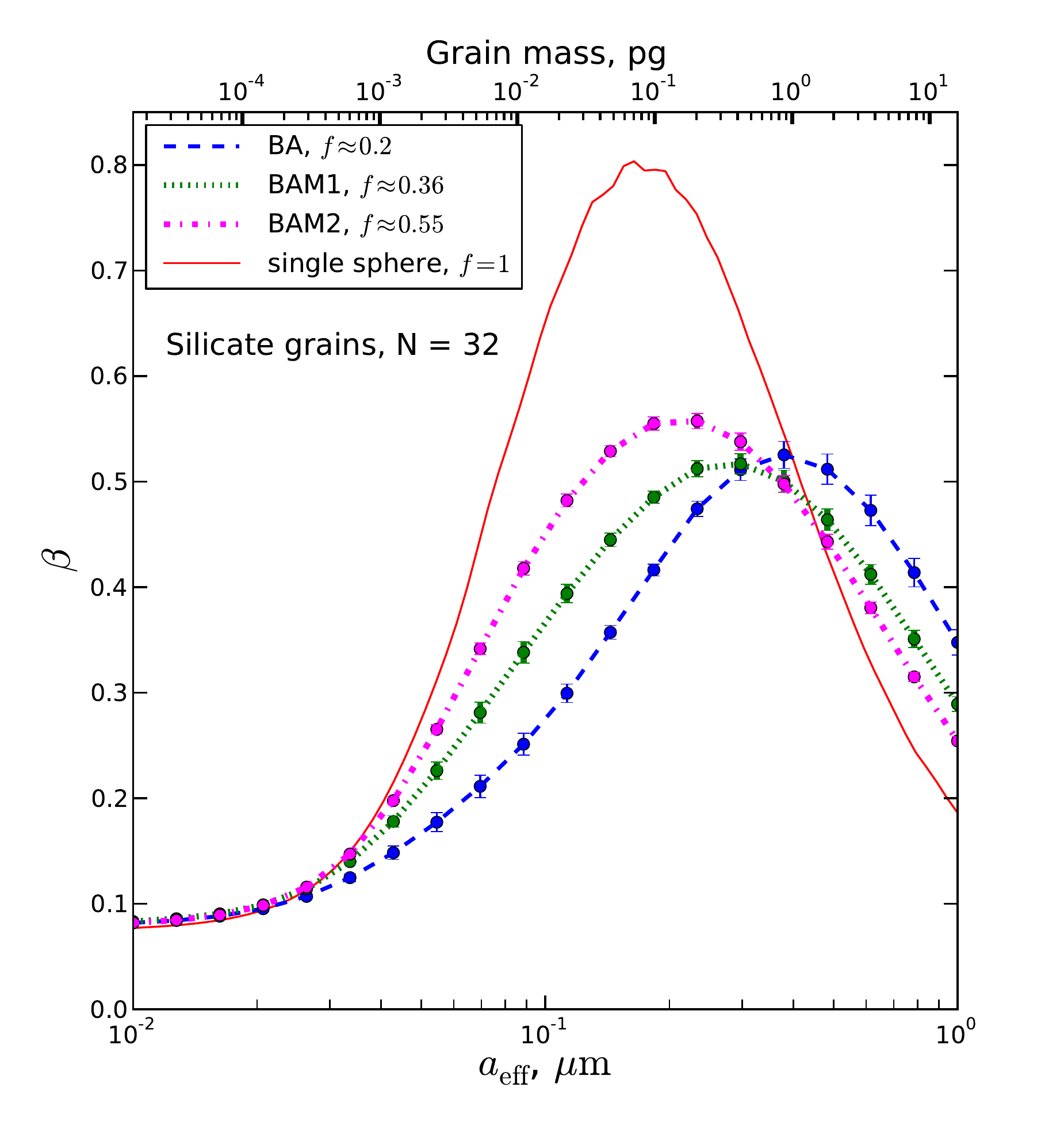}
\caption{$\beta(a_{\rm eff})$ for a solid sphere, and for the three classes of ballistic aggregate targets (section \ref{sect:targets}).   These $\beta$ values are averaged over 27 grain orientations of 16 grains of each class.  Error bars correspond to the standard deviation of the angle-averaged $\beta$ values for each class} 
\label{firstSilicateFigure}
\vspace{-.05cm}
\end{figure}

For the solar spectrum, we present calculations of $\beta(a_{\rm eff})$ for four different sets of geometries.  In order of decreasing compactness, these are: a solid sphere, the BAM2 clusters, the BAM1 clusters, and the BA clusters, all with $N = 32$.  Figure \ref{firstSilicateFigure} shows $\beta(a_{\rm eff})$.  The $\beta$ values are averaged using 27 different orientations for each of 16 grains.  \kern-.5em  \footnote{We used all 16 realizations of the classes BA.32, BAM1.32, and BAM2.32, available at \url{http://www.astro.princeton.edu/~draine/agglom.html}}
\par
We see in Figure \ref{firstSilicateFigure} that fluffiness (low values of $f$) mildly enhances $\beta$ for $a_{\rm eff} \gtrsim 0.3 \mu$m particles, but suppresses $\beta$ for $a_{\rm eff}\lesssim0.3 \mu$m particles.   We also note that for all models, the peak of $\beta$ is less than unity.  All the models converge to within a few percent in the electric dipole limit ($a_{\rm eff} \rightarrow 0$).  This is in contrast to the work of \citet{Kohler07}, which shows grain compactness to have very little effect on $\beta$ for silicate grains between $10^{-2}$ and 10 pg.  Our results are in qualitative agreement with \citet{Tazaki15} who also find that that their less compact grains have lower $\beta$ values except for sizes $\gtrsim 0.5 \mu m$.
\par
The captured {\it Stardust} particles have $M \approx 3$ pg, corresponding to $a_{\rm eff} \approx 0.6 \mu$m, assuming a material density of 3.8 g/cc.  \citet{Westphal14} additionally identify impact craters on the aluminum foil which are consistent with much smaller particles:  $a_{\rm eff} = 0.1-0.15 \mu$m.  Both $a_{\rm eff} = 0.15 \mu$m and $a_{\rm eff} = 0.6 \mu$m are off the peak in Figure \ref{firstSilicateFigure}, reducing the predicted $\beta$ for these sizes to $\lesssim 0.5$.  
\par
 The estimated densities for the two captured {\it Stardust} grains were 0.7 and 0.4 g/cc, corresponding to filling factors $0.1 \lesssim f \lesssim 0.2$, corresponding most closely to the BA aggregates (see Table \ref{table:fillingFactors}).

\begin{figure}
\centering
\includegraphics[scale = .5]{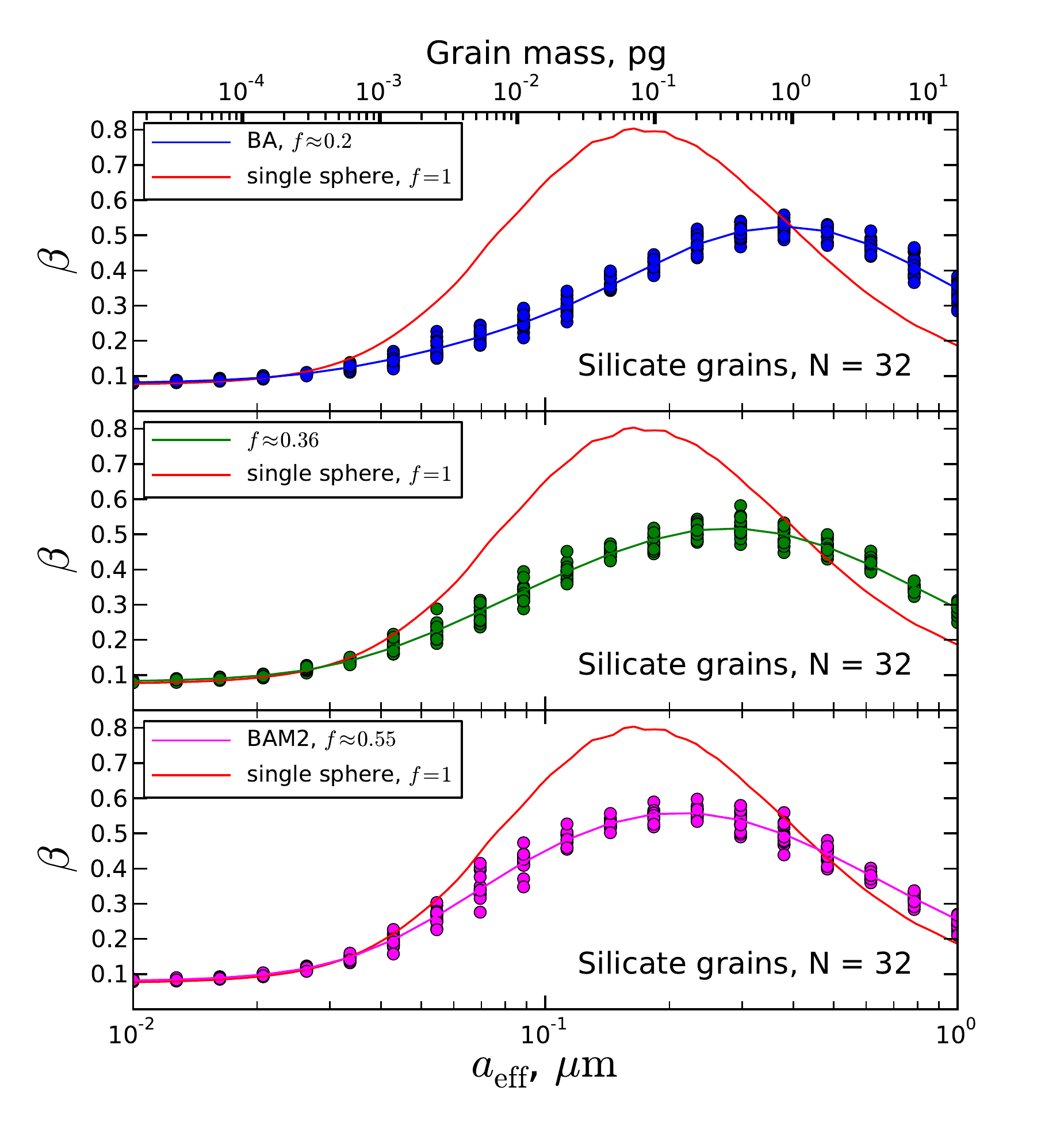}
\caption{Scatter between different targets and orientations.  Each point represents one of the 16 targets for each class at a randomly selected orientation.  Solid curves (identical to those in Figure \ref{firstSilicateFigure}) how the average for each class (BA, BAM1, BAM2) as a function of $a_{\rm eff}$.  In each panel, we have also plotted the value for a single sphere.  }
\label{secondSilicateFig}
\vspace{-.05cm}
\end{figure}
\subsection{Variation of $\beta$ with Grain Realization and Orientation}
\label{sect:orient}
The results presented in the previous section suggest that porous silicate grains  have $\beta$ values that peak below 0.6.  However, there is some scatter arising both from different instances of each class of grain, and also from different possible orientations.  To investigate scatter from these two sources, for each grain and each size, we pick one random orientation, and calculate the $\beta$ value for that grain, size, and angle.  These points are all plotted in Figure \ref{secondSilicateFig}, with the curves (imported from Figure \ref{firstSilicateFigure}) showing the average values for each class.  This gives a sense of the range of likely values for $\beta$ that we would see in a distribution of different grains which enter the solar system with different orientations.  The distribution shown here has a somewhat larger spread than the actual one, as there will be additional averaging over orientations inherent to whatever spin state the grains have.  We see a scatter on the order of 10\% within each class of grain.  All of our conclusions are robust to arbitrary choices of grain instance and orientation angle.
\par
From Figure \ref{secondSilicateFig} we cannot tell if the majority of scatter is due to variations between different grains of the same class, or due to different orientations.  To investigate this, for each grain, we calculate $\beta$ for $N_{\rm ori} = 10$ random orientations.   Then for each grain class and size, we define two statistics 
\begin{equation}
\bar \sigma_{\theta}  \equiv \sum_{i =1}^{N_{\rm gr}} \frac{1}{N_{\rm gr}}  \sqrt{\sum_{j=1}^{N_{\rm ori}}\frac{(\beta_{i, j} - \mu_{i})^2}{N_{\rm ori} - 1}}, 
\end{equation}
and 
\begin{equation}
\sigma_{\rm tot} \equiv \sqrt{\sum_{i=1}^{N_{\rm gr}}  \sum_{j=1}^{N_{\rm ori}} \frac{(\beta_{i, j} - \mu)^2}{N_{\rm ori}N_{\rm gr} - 1}}, 
\end{equation}
where $\mu_i$ is the value of $\beta$ averaged over orientations of a single grain, and $\mu$ is the value of $\beta$ averaged over all grains and orientations.    $\bar \sigma_\theta $ is the mean standard deviation in $\beta$ for a given grain sampled at different orientations.  $\sigma_{\rm tot}$ is the standard deviation in $\beta$ for a given {\it grain class} sampled for different instances and orientations.  If most of the variance is due to different grain instances, rather than orientations, then we expect $\bar \sigma_{\theta} \ll \sigma_{\rm tot}$.  These quantities are tabulated for two different grain sizes.  In the last column we compare the variance attributable to orientation with the total variance within each grain class, and come to the conclusion that the majority of variation in $\beta$ is due to grain orientation, rather than grain instance.

\begin{table}[ht]
\caption{Scatter in $\beta$ due to angles and instances}
\centering
\begin{tabular}{c c c c}
\hline\hline
Grain Class & $\sigma_{\rm tot}$ & $\bar \sigma_{\theta} $ & $(\frac{\bar \sigma_{\theta}}{\sigma_{\rm tot}})^2$\\ [0.5ex] 
\hline 
BA 0.15 $\mu$m & 0.0181 & 0.0156 & 0.743 \\ 
BAM1 0.15 $\mu$m &  0.0161 & 0.0153 & 0.903 \\ 
BAM2 0.15 $\mu$m &  0.0137 & 0.0117 & 0.729 \\ 
BA 0.6 $\mu$m &  0.0236 & 0.0201 &   0.725 \\ 
BAM1 0.6 $\mu$m  &  0.0187 & 0.0169 & 0.817 \\ 
BAM2 0.6 $\mu$m &  0.0182 & 0.0170 & 0.872 \\ [1ex]
\hline
\end{tabular}
\label{table:scatterCauses}
\end{table}

\begin{figure}
\centering
\includegraphics[scale = .5]{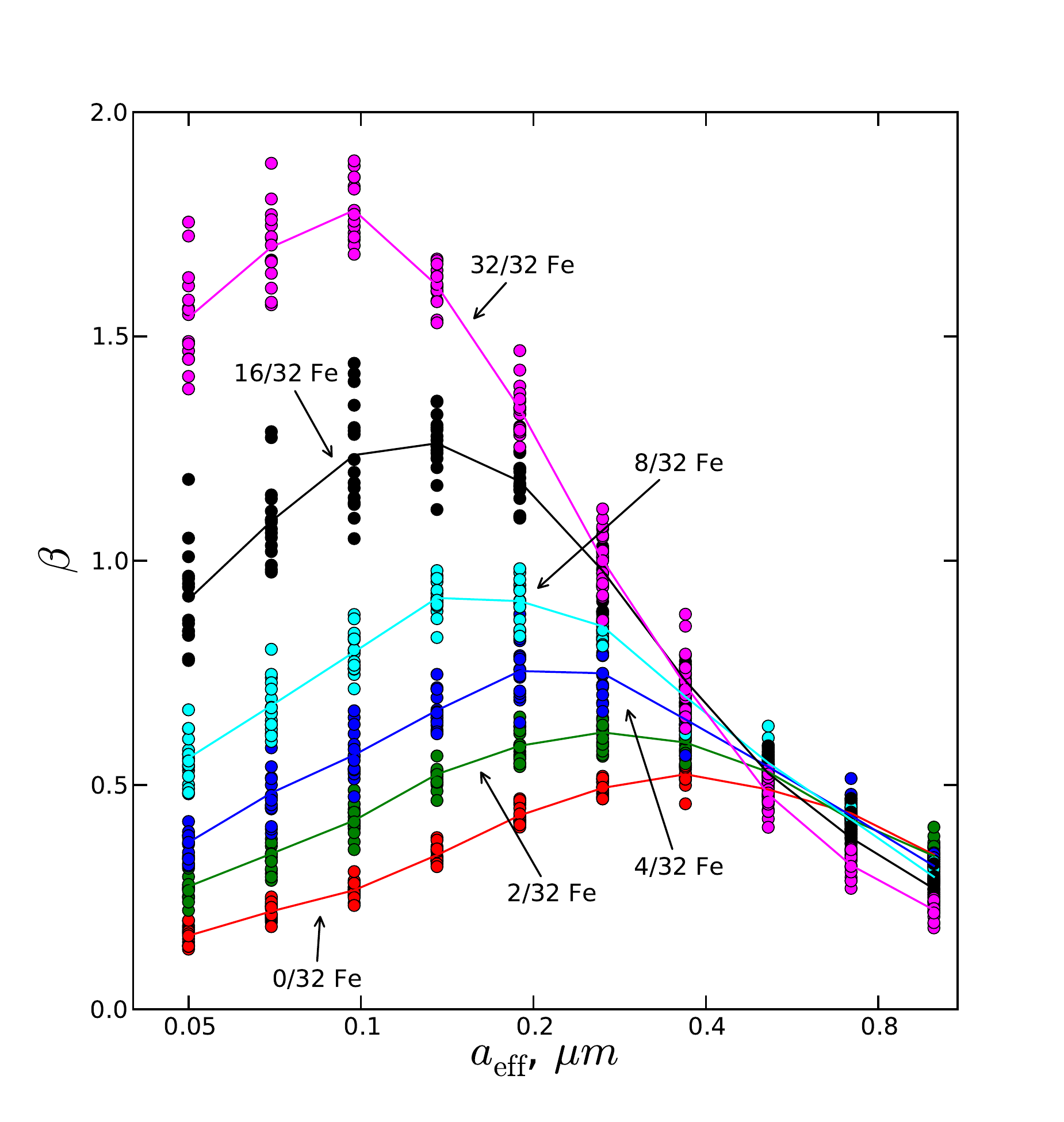}
\caption{ Curves of $\beta$ vs. $a_{\rm eff}$ for targets of the BA class (BA.32.1-BA.32.16), with random orientations.  This is done for targets where some of the 32 silicate spheres have been replaced with iron spheres of the same size.  Solid lines represent averages of all the 16 data points. }
\label{ironFig}
\vspace{-.05cm}
\end{figure}

\subsection{Metallic Iron Inclusions}
\label{iron}
The {\it Stardust} team reports \citep{Westphal14} the presence of ``Fe-bearing phases" in two captured grains (``Orion" and ``Hylabrook"), possibly consistent  with metallic iron.  Some interstellar grains may contain a substantial amount of metallic iron \citep[see][]{Jones90, Draine13}.  We calculate the effect of Fe on $\beta$ by replacing some of the spheres in the target structures with material having the density and dielectric function of iron, rather than astrosilicate.  This was done assuming that the iron spheres are randomly located (that is,  for a grain with $N$ spheres, $M$ of which are iron, any of the $N \choose M$ configurations are equally likely to be chosen).  Figure \ref{ironFig} shows a scatterplot of $\beta$ values for different numbers of iron spheres and different effective radii for 16 members of the BA class of targets.  For each target, we picked a random orientation.  We see that the addition of iron substantially increases $\beta$ for $a_{\rm eff} \lesssim 0.3 \mu$m.  However, to make $\beta$ exceed one, we require $\gtrsim 35\%$ of the solid volume be iron,  and $a_{\rm eff} \lesssim 0.25\, \mu$m.  Such grains have $<0.16$ pg of silicate material, far below the $\sim 3-4$ pg masses of Orion and Hylabrook. 
\subsection{Carbon}
The captured grains might possibly have had carbonaceous mantles, perhaps lost during the collection process.  We consider three different experimentally determined dielectric functions for amorphous carbon from \citet{Zubko96}, and show that we cannot bring $\beta$ above unity for a grain with 3 pg of silicate material by adding carbon to the grains.
\par

\begin{figure}
\centering
\includegraphics[scale = .4]{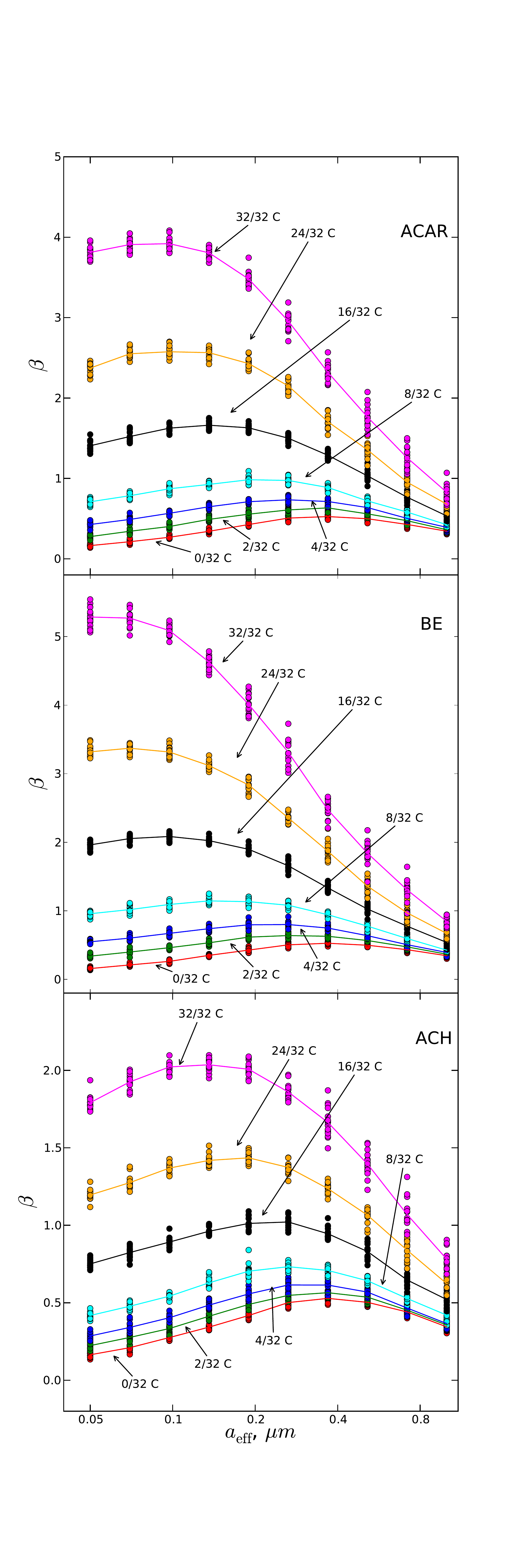}
\caption{Each panel of this figure is analogous to Figure \ref{ironFig}, for a different carbon dielectric function.}
\label{carbon1}
\vspace{-.05cm}
\end{figure}

\begin{figure}
\centering
\includegraphics[scale = .5]{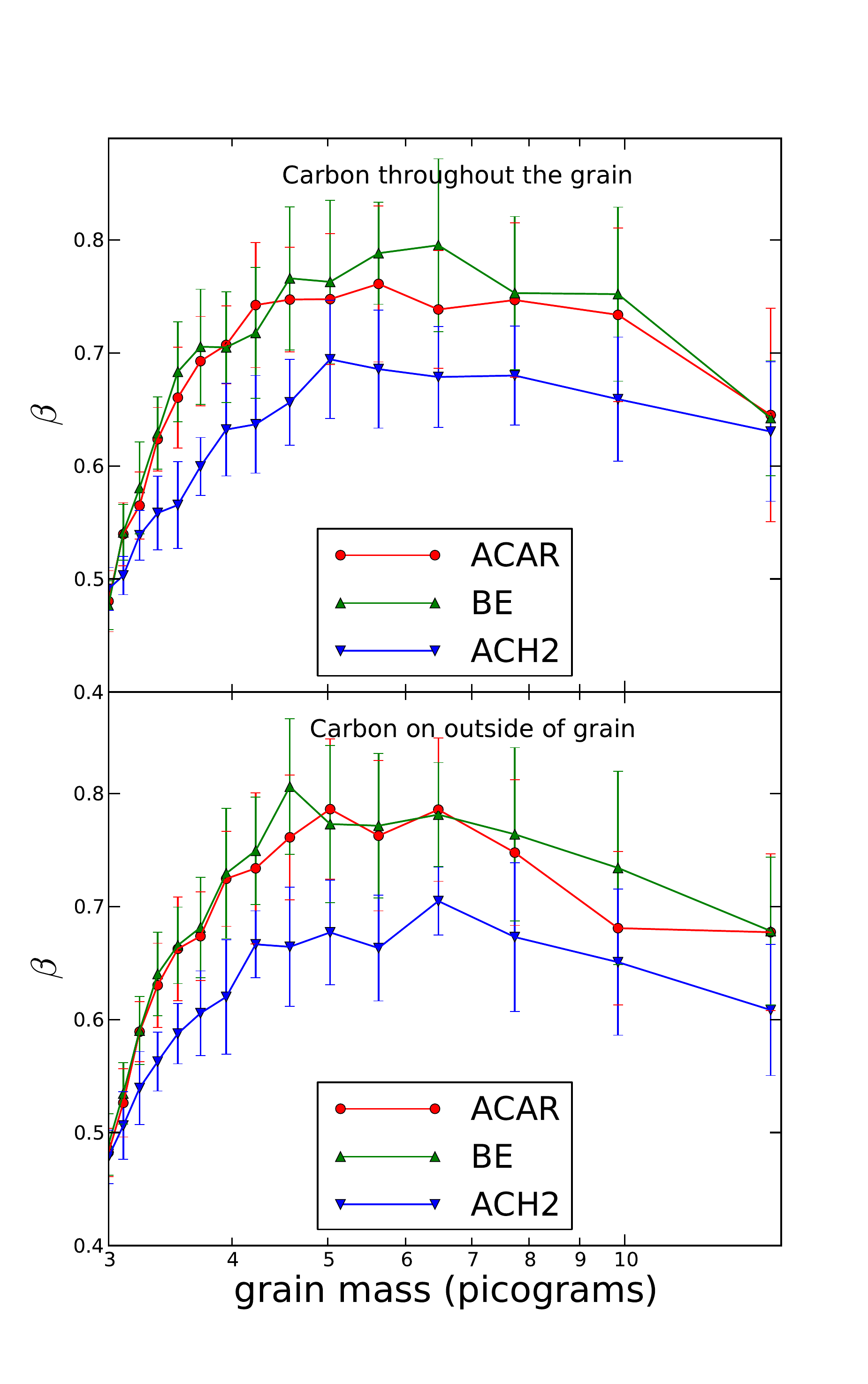}
\caption{$\beta$ as a function of total mass for grains which contain 3 picograms of silicate material.  The upper panel shows grains with $N$ = 32 spheres, with carbonaceous spheres distributed randomly.  The lower panel shows grains with $N = 32$ spheres, where the silicate spheres arrive first during the formation process, and the carbon spheres arrive later, forming a fluffy ``mantle".  Error bars represent the $1-\sigma$ scatter from grain to grain.}
\label{carbon2}
\vspace{-.05cm}
\end{figure}

Figure \ref{carbon1} shows $\beta$ as a function of $a_{\rm eff}$ for different values of the carbon fraction.  Carbon spheres were distributed randomly throughout the grain as in Section \ref{iron}.  We assume a carbon density of 2.0 g cm$^{-3}$ \citep{Robertson86}.  We see that $\beta$ peaks above 1 for all three dielectric functions if 1/4 or more of the material by volume is carbon.  However, $\beta > 1$ occurs only for grains with small silicate masses.  Figure \ref{carbon2} shows $\beta$ for particles containing 3 picograms of silicate material \citep[motivated by the 3 putative interstellar particles captured by {\it Stardust}][]{Westphal14}, as a function of total grain mass, assuming the non-silicate mass to be in the form of amorphous carbon.  The top panel assumes the carbon spheres to be distributed randomly throughout the grain, and the bottom panel assumes that the spheres which impact the structure later during the generation algorithm are carbon (leading to carbon preferentially on the outside of the grain).  In no case do we find $\beta > 1$ for grains with 3 pg of silicate material.

 \subsection{Number of Spheres}
\begin{figure}
\centering
\includegraphics[scale = .54]{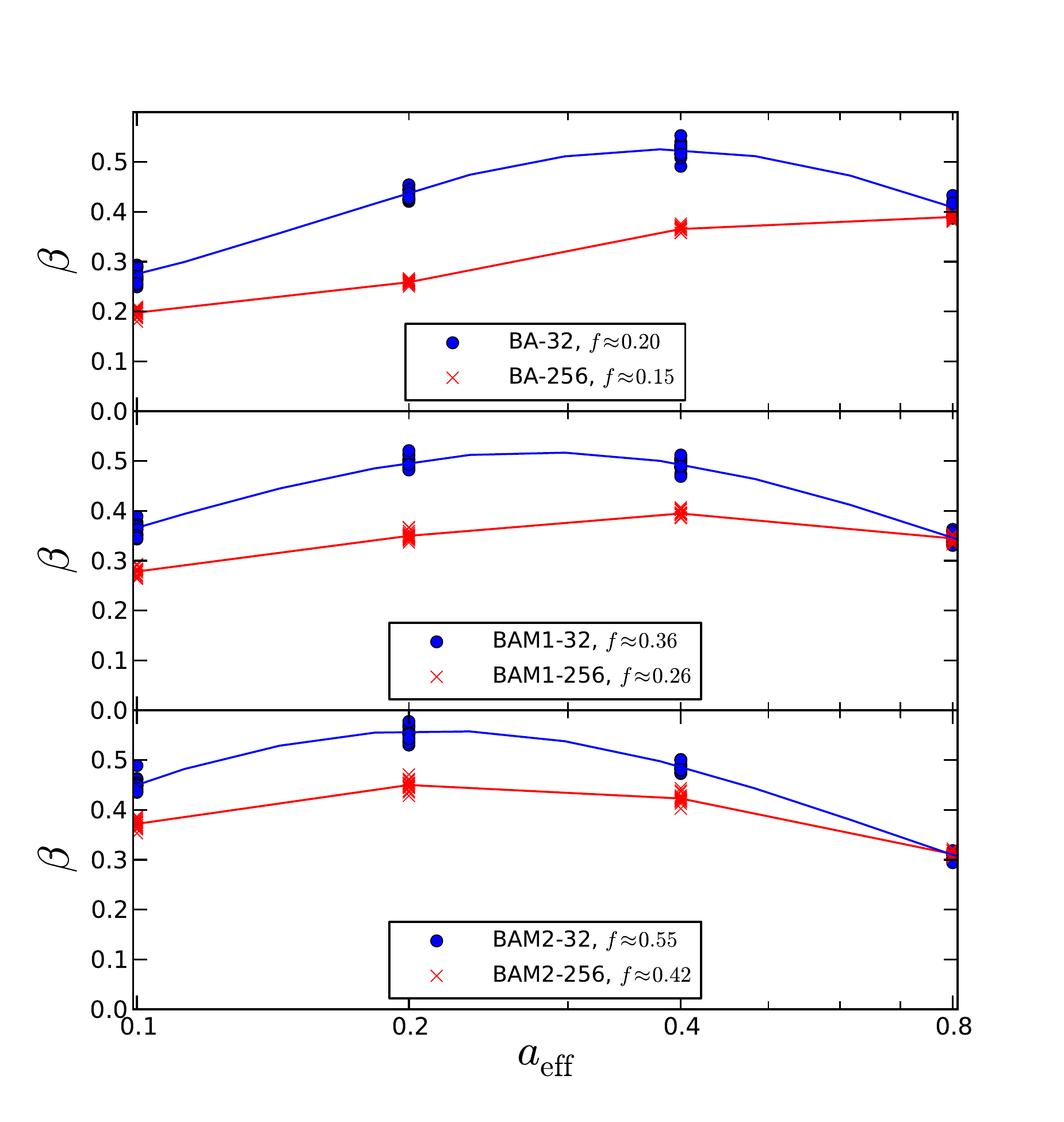}
\caption{Variation of $\beta$ depending on number of spheres in the target cluster.   Each point represents one of the 16 targets for each class at a randomly selected (from an isotropic distribution) angle.  Each panel represents a different grain model, as indicated.  Instances with $N = 32$ are represented with filled circles.  Instances with $N = 256$, are represented by crosses.  Solid lines represent average values.  These are taken from Figure \ref{firstSilicateFigure} for the case with 32 spheres, and calculated as the mean of the data points for the 256 spheres case.}
\label{256Fig}
\vspace{-.05cm}
\end{figure}

 We also consider changing the number of spheres in our model dust grain.  We consider model grains generated by the same algorithm, but with $N = 256$ (the largest $N$ for which the calculations remain tractable) instead of 32.  Figure \ref{256Fig} is analogous to Figure \ref{secondSilicateFig}: for each member of each target class, we pick a random orientation, and calculate $\beta$.  We see that there can be substantial differences between $N = 32$ and $N = 256$ (the filling factor drops by $\sim 20 \%$ as $N$ increases from 32 to 256), particularly in the value of $a_{\rm eff}$ where $\beta$ peaks, but our principal conclusion \textemdash \; that $\beta$ peaks substantially below unity for silicate clusters \textemdash \; remains robust regardless of the number of spheres used.  
\newpage

\section{transverse forces}
\label{transForce}
When modeling the dynamics of dust grains entering the solar system, it may not be appropriate to ignore the transverse force from radiation pressure \citep{Kimura02}.  It is difficult to correctly calculate the transverse force averaged over the rotation of the grain, as the spin state depends on the initial spin state far from the Sun, torques from solar radiation pressure, and internal relaxation processes.  Here, we present an estimate of the impulse delivered to the grain due to transverse radiation forces.  For each of the target geometries, we used the discrete dipole approximation code DDSCAT \citep{Draine94, ddscat}.  Because DDSCAT is slower than the MSTM code, we used $N_\lambda = 10$ instead of 30 to speed up calculations.  Based on the results in the appendix, we expect this to make less than a 2\% difference to the results. 
\par
We calculated values of the ``transverse $\beta$" or $\beta_t$.  This is the ratio of transverse radiation force to gravitational force.  $\beta_t$ will vanish if we average over all grain orientations, so we calculated it for grains spinning about their principal axis of largest moment of inertia, for different values of the angle $\theta_{a1}$ between the spin axis and the direction to the Sun.  This was calculated for $a_{\rm eff} = 0.15$ and 0.6 $\mu$m grains of the BAM1 geometry.  Results are shown in Table \ref{table:betat}.   Because we average over rotations about the spin axis, $\beta_t(\theta_{a1} = 0) = 0$ by symmetry.
\begin{table}[ht]
\caption{Transverse $\beta$ statistics for BAM1, $N$ = 32 (realizations BAM1.32.1 - BAM1.32.16)}
\centering
\begin{tabular}{c c c c}
\hline\hline
$a_{\rm eff}, \mu$m &$\theta_{a1}$ & $\langle \beta_t \rangle$ & max$(\beta_t)$ \\ [0.5ex] 
\hline 
0.15 &$0^\circ$ &  $0$ & $0$ \\ 
0.15 & $30^\circ$ &  0.067 & 0.098  \\ 
0.15 & $60^\circ$ &  0.087 & 0.129  \\ 
0.15 & $90^\circ$ &  0.0051 & 0.0097 \\
0.6 & $0^\circ$ &  $0$ & $0$  \\ 
0.6 & $30^\circ$ &  0.011 & 0.026  \\ 
0.6 & $60^\circ$ &  0.021 & 0.047  \\ 
0.6& $90^\circ$ &  0.0077 & 0.020 \\ [1ex]
\hline
\label{table:betat}
\end{tabular}
\end{table}
\par
We can estimate the transverse velocity arising from radiation pressure by considering an instructive case with $\beta = 1$, and $\beta_t \ll 1$.  This case has straight-line constant-velocity particle trajectories except for a small perturbation due to $\beta_t$.  Let us assume that the transverse force acts in the same direction over the trajectory.  This would be the case if both the transverse force and the grain's spin axis were perpendicular to the orbital plane, but provides a reasonable estimate and an upper bound for more complicated geometries. In this case the total transverse velocity acquired by the grain on its journey from infinity to the point $(R, \theta)$ (where $R$ is the grain-Sun distance, and $\theta$ is the angle between the interstellar wind direction and the grain-Sun vector) is:
\begin{equation}
v_t = \int_{R\cos{\theta}}^\infty \frac{\beta_t GM_\odot dx}{v_i(x^2 + (R\sin{\theta})^2)} = \frac{\beta_t GM_\odot}{Rv_i}\frac{\theta}{\sin{\theta}},
\end{equation}
where $v_i$ is the velocity of the Sun with respect to the local ISM.  For numbers ($R = 2 AU$, $v_i = 26$ km s$^{-1}$ and $\theta = 60^{\circ}$) appropriate for the {\it Stardust} mission \citep{Sterken14}, this gives $v_t = 21 \beta_t$km s$^{-1}$.  Using typical numbers from Table \ref{table:betat}, we see that we expect changes in transverse velocity of $\sim 1$ km s$^{-1}$ for the small grains that left impact craters in the {\it Stardust} mission, and $\sim 0.2$ km s$^{-1}$ for the large captured grains.  Including a true value of $\beta$ less than one should slightly lower the transverse velocity, as the transverse force has less time to act due to the increased radial velocity.  
\section{Torques}
\label{torquesec}
We can also compute radiative torques on dust grains using the DDSCAT code.  We computed the component of these torques along the spin axis for the `BAM1' class of grains, with effective radii of 0.15 and 0.6 $\mu$m.  As in the previous section we calculated torques for grains spinning about their principal axis of largest moment of inertia, for different values of the angle $\theta_{a1}$ between the spin axis and the grain-Sun vector. We evaluate the torque using DDSCAT.  The torque efficiency $\vec Q_{\Gamma}(\lambda)$ is 
\begin{equation}
\vec Q_{\Gamma} \equiv \frac{2\pi \vec \Gamma}{\pi a_{\rm eff}^2 u_{\rm rad} \lambda}
\end{equation}
\citep{DW96}, where $\vec \Gamma$ is the torque, and $u_{\rm rad}$ the energy density of radiation with wavelength $\lambda$.  Given a spectrum of flux $F(\lambda)$, the torque is
\begin{equation}
\vec \Gamma = \pi a_{\rm eff}^2 u_{\rm rad} \frac{\langle \lambda \rangle}{2 \pi} \langle \vec Q_\Gamma \rangle,
\end{equation}
where $u_{\rm rad}$ is the energy density of the incident radiation,
\begin{equation}
\langle \vec Q_\Gamma \rangle = \frac{\int_0^\infty  \lambda F(\lambda) \vec Q_\Gamma d \lambda}{\int_0^\infty \lambda F(\lambda) d\lambda},
\end{equation}
and
\begin{equation}
\langle \lambda \rangle \equiv  \frac{\int_0^\infty \lambda F(\lambda) d \lambda}{\int_0^\infty F(\lambda) d\lambda}.
\end{equation}
Assuming the trajectory to be undeflected by the combination of radiation pressure forces and solar gravity (i.e. $\beta = 1$, $\beta_t = 0$), we can integrate $\Gamma(r) dt = \Gamma(r) dr/v_r(r)$ along the grain trajectory (analogous to the calculation in Section \ref{transForce}), assuming zero spin angular momentum at infinity, to find the spin angular momentum of the grain:
\begin{equation}
\label{eq:Lbig}
\vec L(R) = \frac{\langle \vec Q_\Gamma \rangle a_{\rm eff}^2L_\odot \langle \lambda \rangle}{8\pi v_i R c}\frac{\theta}{\sin{\theta}}.
\end{equation}
\par
This calculation neglects rotational damping due to emission of infrared photons.  The damping torque on a spinning grain from radiation of photons with wavelengths long compared with the size of the grain is given by \citep{AHD09}
\begin{equation}
\frac{dL}{dt} = \frac{-\omega}{2 \pi^2} \int_0^{\infty}\frac{P_\nu d\nu}{\nu^2},
\end{equation}
where $P_{\nu}$ is the radiated power per unit frequency.  Let the grain have temperature $T_{\rm gr} = \gamma T_{\rm BB}$, where $T_{\rm BB} = T_\odot (0.5R_{\odot}/R)$ is the ``blackbody" temperature.  Submicron grains are poor thermal radiators, and will likely have $\gamma > 1$.  For an infrared opacity $\propto \nu^{s}$, the damping torque is 
\begin{equation}
\frac{dL}{dt} = \frac{-2}{\pi \gamma^2} \; \; \frac{\Gamma(2+s) \zeta (2+s)}{\Gamma(4+s) \zeta(4+s)} \frac{h^2}{k^2} \sigma T_{\rm BB}^2 \langle Q_{\rm abs} \rangle a_{\rm eff}^2 \omega,
\end{equation}
where $\langle Q_{\rm abs} \rangle$ is the absorption efficiency averaged over the solar spectrum, and $\zeta(x)$ is the Riemann $\zeta$ function.  For $s = 2$, this becomes 
\begin{equation}
\frac{dL}{dt} = -0.034\; \frac{h^2}{k^2 \gamma^2} \; \sigma T_{\rm BB}^2 \langle Q_{\rm abs} \rangle a_{\rm eff}^2 \omega.
\end{equation}
Since 
\begin{equation}
L = \frac{8 \pi}{15} \rho a_{\rm eff}^5 \omega f^{2/3},
\end{equation}
the damping timescale $\tau = L/\dot L$ is 
\begin{align}
&& \tau = \frac{8 \pi \rho a_{\rm eff}^3 k^2 \gamma^2}{15 \times 0.034 h^2 \sigma T_{\rm BB}^2 \langle Q_{\rm abs} \rangle f^{2/3}} & \nonumber\\
&&= \frac{27 {\rm yr}}{\langle Q_{\rm abs} \rangle} \gamma^2 \left(\frac{a_{\rm eff}}{0.2 \mu {\rm m}}\right)^3 \left(\frac{\rm R}{2 \rm AU} \right)^2 \left(\frac{0.2}{f}\right)^{2/3}.
\end{align}
\par
Given the expectation that $\gamma > 1$, this is much longer than the timescale on which grains gain their spin (which is just the dynamical time $R/v \sim 0.37 R/(2{\rm AU})$ years, since torque is proportional to the gravitational acceleration of the grain, so $\vec L$ should vary on the same timescale as $v$) for 0.15-0.6 $\mu$m grains at 2 AU, so we can ignore spin damping.
\par

Using Equation \eqref{eq:Lbig} and letting $\theta/\sin{\theta} \approx 1$, Table \ref{table:surfaceVelocities} shows the mean and maximum surface velocity $v_{\rm surf} \approx L a_{\rm eff}f^{-1/3}/I$ for $a_{\rm eff} = 0.15 \mu$m and  0.6 $\mu$m grains as a function of $\theta_{a1}$, the angle between the rotation axis and the direction to the star.  These are calculated from an ensemble of 16 grains of each size.
\begin{table}[ht]
\caption{Surface velocity for interstellar BAM1 grains at $R \approx 2$ AU}
\centering
\begin{tabular}{c c c c}
\hline\hline
$a_{\rm eff}, \mu$m & $\theta_{a1}$, degrees & mean $v_{\rm surf}$, m s$^{-1}$  & max $v_{\rm surf}$, m, m s$^{-1}$ \\ [0.5ex] 
0.15 & 0 & 458 & 945 \\
0.15 & 30 & 368 & 727 \\
0.15 & 60 & 205 & 495 \\
0.15 & 90 & 74 & 226 \\
0.6 & 0 & 471 & 1473 \\
0.6 & 30 & 230 & 560 \\
0.6 & 60 & 72 & 197 \\
0.6 & 90 & 115 & 241 \\ [1ex]
\hline
\end{tabular}
\label{table:surfaceVelocities}
\end{table}
\par
Having estimated surface velocities, we then ask whether this will lead to centrifugal disruption.  We can estimate the critical velocity by equating the centrifugal force pulling two hemispheres apart with the cross-sectional area times the yield stress $S_{\rm max}$.  This calculation shows that centrifugal disruption will occur for $v \gtrsim v_{\rm crit} = 2\sqrt{S_{\rm max}/\rho}$.  For ideal materials with no fractures, $S_{\rm max} \sim 10^{11}$ dyn cm$^{-2}$ \citep{MacMillan72}, and $v_{\rm crit}$ is of order the sound speed in the material. 
\par
 However, because of defects, real materials have tensile strengths well below the ideal value.  If we use a density of 2.4 g cm$^{-3}$, and $S_{\rm max} = 5.0\cdot 10^{7}$dyn cm$^{-2}$,  appropriate for construction grade concrete \citep{Anoglu06}, we obtain $v_{\rm crit} \approx 90$ m s$^{-1}$.  This is smaller than the estimates for $v_{\rm surf}$ in Table \ref{table:surfaceVelocities}, implying that some grains may be centrifugally disrupted by the solar radiation torques. 
 \par 
The spin-up of submicron grains by solar radiation is analogous to the YORP (Yarkovsky-O'Keefe-Radzievskii-Paddack) effect (see \citet{Vokrouhlicky15} for a review), but where the irregularities are on scales comparable to the wavelength.  Dust grains have far higher critical angular velocities for disruption than rubble piles because the former are held together by chemical bonds rather than gravity.
\section{Particles Captured by {\it Stardust}}
The {\it Stardust} mission captured two particles (``Orion" and ``Hylabrook") that are claimed to be consistent with interstellar dust particles entering the solar system.  The condition of these particles is consistent with impact velocities $\ll \!10$ km s$^{-1}$ \citep{Westphal14}.  The particle masses correspond to $a_{\rm eff} \approx 0.6 \mu$m, assuming a material density $\sim \!3.8$ g cm$^{-3}$.  
\par
Impact speeds $<\!$10 kms$^{-1}$ require $\beta > 1$, so that solar radiation pressure can decelerate incoming grains (see Figure 8 of \citet{Sterken14}).  Using silicate grains, supplemented with carbon and iron, we are unable to find a composition such that $\beta > 0.9$ for a grain with 3 picograms of silicate material (see Figures \ref{ironFig} and \ref{carbon2}).  Therefore, if the actual impact speeds are indeed $<10\!$km s$^{-1}$, it seems unlikely that particles Orion and Hylabrook originate in the interstellar stream approaching the solar system at $\sim \! 26$ km s$^{-1}$; some other origin, presumably in the solar system, appears to be required.  While we have not considered every possible set of grain properties, we have covered much ground, both in composition and structure, without managing to produce a grain with high enough $\beta$.
\section{Summary}

\begin{enumerate}
\item Accurate calculations of models for ``fluffy" grains with the dielectric functions and density appropriate for astronomical silicates show that $\beta < 1$ for all sizes.  High grain porosity does not enable sub-micron sized grains to have substantially higher ratios of radiation pressure to mass.  In fact, for $a_{\rm eff} = 0.1 - 0.15 \mu$m grains (such as those responsible for the {\it Stardust} Al foil impact craters), higher porosity tends to substantially reduce $\beta$ at constant $a_{\rm eff}$.

\item One way to potentially increase $\beta$ is to include metallic iron.  However, we find that $\beta$ peaks below one unless about half the mass of the grain is iron.  For the grains with the masses of those captured intact in aerogel by the {\it Stardust} mission, $\beta$ would be $\lesssim 0.6$ even if the grain were entirely iron. 

\item We also consider mixed silicate-amorphous carbon structures.  For grains with $\sim \!$3pg silicate mass (as for the Orion and Hylabrook grains captured by {\it Stardust}) we find $\beta \lesssim 0.8$ (see Fig \ref{carbon2}).  For particles with mass and composition resembling the captured {\it Stardust} grains, we are unable to find a single example with large enough $\beta$ for the particle to be of interstellar origin, even if they are allowed to have amorphous carbon mantles that were lost during capture.

\item Transverse forces from radiation pressure on incoming interstellar grains lead to velocity changes of at most $\sim \!2$ km s$^{-1}$ at 2 AU. 

\item Radiative torques due to sunlight can drive irregular sub-micron grains entering the solar system to spin with equatorial velocities of a few hundred meters per second.  Depending on the tensile strength of the grains, this could lead to rotational break-up.

\end{enumerate}

\section{Acknowledgments}
We are grateful to D. Mackowski for the availability of the MSTM code.  We thank the referee for helpful suggestions that led to improvements in the paper.
\\
B. T. D. was supported in part by NSF grant AST 1408723.
\bibliographystyle{apj}
\bibliography{apj-jour,radiationPressureDraft}
\appendix

\begin{figure}
\centering
\includegraphics[scale = .5]{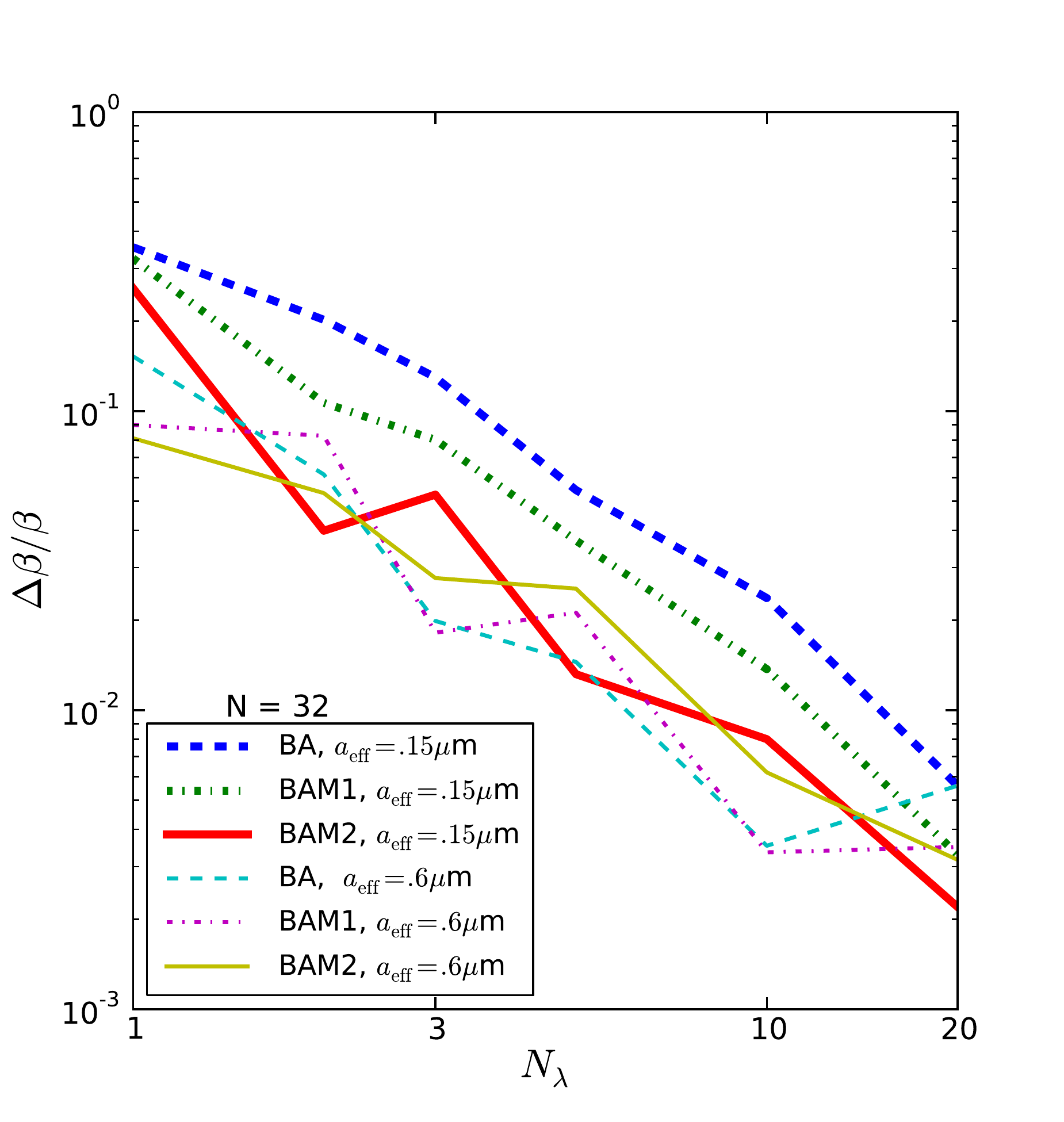}
\caption{Magnitude of fractional change in $\beta$ when using number of wavelengths $N_\lambda < 30$.  Errors are under 5\% relative to $N_\lambda = 30$ for values of $N_\lambda$ as low as 5.}
\label{deviation}
\vspace{-.05cm}
\end{figure}

We investigate the sensitivity of the calculated $\beta$ to the choice of $N_\lambda$, the number of wavelengths sampled.  Figure \ref{deviation} shows the magnitude of the fractional change in $\beta$, relative to the result for $N_\lambda = 30$, for different grains as labelled on the plot.  This is averaged over one orientation of 16 different grains in each class.  We see very good convergence at the few percent level for all $N_\lambda > 5$.  
\end{document}